\documentclass[a4paper]{article}
\usepackage{ISCSLP2024}
\usepackage{graphicx} 		
\usepackage{subfigure}		
\usepackage{CJKutf8}
\usepackage{hyperref}
\usepackage{amsmath}
\usepackage{ifthen}
\newboolean{blind}
\setboolean{blind}{False} 
\title{The Codec Language Model-based Zero-Shot Spontaneous Style TTS System for CoVoC Challenge 2024}
\name{
	\ifthenelse{\boolean{blind}}{Anonymous to ISCSLP}
	{Shuoyi Zhou, Yixuan Zhou, Weiqin Li, Jun Chen, \\
 Runchuan Ye, Weihao Wu, Zijian Lin, Shun Lei, Zhiyong Wu$^{*}$\thanks{$^{*}$ Corresponding author.}}
}
\address{
  \ifthenelse{\boolean{blind}}{Anonymous to ISCSLP}
  {
  	Shenzhen International Graduate School, Tsinghua University, Shenzhen
  }
}
\email{
	\ifthenelse{\boolean{blind}}{Anonymous to ISCSLP}
	{
        zhousy23$@$mails.tsinghua.edu.cn, zywu$@$sz.tsinghua.edu.cn
        }
}

\begin{document}

\maketitle
\begin{abstract}
  This paper describes the zero-shot spontaneous style TTS system for the ISCSLP 2024 Conversational Voice Clone Challenge (CoVoC). 
  We propose a LLaMA-based codec language model with a delay pattern to achieve spontaneous style voice cloning. 
  To improve speech intelligibility, we introduce the Classifier-Free Guidance (CFG) strategy in the language model to strengthen conditional guidance on token prediction. 
  To generate high-quality utterances, we adopt effective data preprocessing operations and fine-tune our model with selected high-quality spontaneous speech data.
  The official evaluations in the CoVoC constrained track show that our system achieves the best speech naturalness MOS of 3.80 and obtains considerable speech quality and speaker similarity results.
\end{abstract}

\section{Introduction}


 Text-to-speech (TTS) aims to create natural and human-like speech for given input text. 
 On the one hand, TTS systems need to accurately replicate the target speakers' voice, including their timbre, pitch, and prosody, using voice cloning techniques. 
 On the other hand, these systems must be able to model spontaneous speech elements like pauses, hesitations, coughs, laughter, and breathing. 
 As a result, voice cloning and spontaneous style modeling are becoming increasingly important areas of focus.


Traditional approaches \cite{ping2018deep} used high-qualtiy $\langle$text, speech$\rangle$ paired data with speaker identity labels to train a
multi-speaker TTS system,
but they could only generate voices of speakers present in the training dataset. 
Later methods \cite{arik2018neural, chenadaspeech}, which were improved by fine-tuning pre-trained multi-speaker TTS models with a few samples of new speakers (called TTS adaptation), still required additional training iterations and model parameters.
Inspired by advances in large language models, recent TTS systems \cite{wang2023neural,zhang2023speak,wang2024speechx,song2024ella} tended to leverage a prompt-based language model (LM) paradigm for this task.
Toward this, speech data was encoded by
neural audio codec models \cite{defossez2022high,zeghidour2021soundstream} into discrete tokens for LM. 
The LM-based TTS models could be trained on large, diverse, and low-quality speech datasets without the need of speaker identity labels.
Benefiting from large-scale training data and LM's in-context learning abilities, 
these models showcased powerful zero-shot voice cloning capabilities, transcending the constraints of  traditional methods.

Although synthesized speech has achieved certain effects in terms of timbre similarity, 
it performs inadequately in daily conversational scenarios. 
The primary issue lies in the insufficient modeling of spontaneous style speech, which encompasses a wealth of paralinguistic information.
Some researchers explored spontaneous speech synthesis by training neural TTS models on high-quality and well-annotated spontaneous speech corpora \cite{guo2021conversational}.
Modeling global and local speaking style representations was proven to be effective in enhancing the naturalness of spontaneous prosody \cite{mitsui2022end,li2024spontts}.
Additionally, explicitly modeling spontaneous behaviors could also introduce more natural paralinguistic phenomena in synthesized speech \cite{cong2021controllable}.
Yet, due to the scarcity of high-quality spontaneous speech data and limitations in model capabilities, these methods still exhibited noticeable disparities in the naturalness of synthesized speech compared to real human speech.
Recently, SponLMTTS \cite{li2024spontaneous} classified spontaneous behaviors into 19 categories. 
It proposed a speech synthesis scheme using LM as the model backbone. 
This scheme involved pre-training on large-scale data followed by fine-tuning with high-quality spontaneous style speech data, achieving breakthroughs in spontaneous style speech synthesis.
However, SponLMTTS focused only on modeling spontaneous styles, and it  ignored voice cloning due to the absence of multi-speaker spontaneous speech data.

The Conversational Voice Clone Challenge (CoVoC), as one of the grand challenges in ISCSLP 2024, aims to promote the development of expressive and spontaneous speech synthesis technology in zero-shot scenarios. 
Our team participated in the constrained track, where we can use the official provided challenge dataset to train a TTS model for spontaneous style voice cloning.
To achieve this, we have developed a codec language model-based TTS model with a delay pattern to support spontaneous style voice cloning.
The delay pattern's autoregressive (AR) structure is more suitable for modeling the rich prosodic variations found in spontaneous speech.
To improve intelligibility, we incorporate Classifier-Free Guidance (CFG) strategy into the codec language model, which can strengthen conditional guidance on token prediction during inference.
For data preprocessing, we slice long sentence, denoise by removing non-speech sounds and classify noise levels.
For model training, we pre-train our model with large-scale data to achieve basic speech synthesis and voice cloning.
We then fine-tune our model with high-quality and spontaneous multi-speaker data, allowing the model to learn spontaneous speech styles from the data.
According to the official evaluations, our system is ranked 3rd in the constrained track and achieves the best speech naturalness MOS of 3.80 and obtains considerable speech quality and speaker similarity results.


\vspace*{-\topskip}
\begin{figure*}[!ht]
  \centering
  \includegraphics[width=0.7\textwidth]{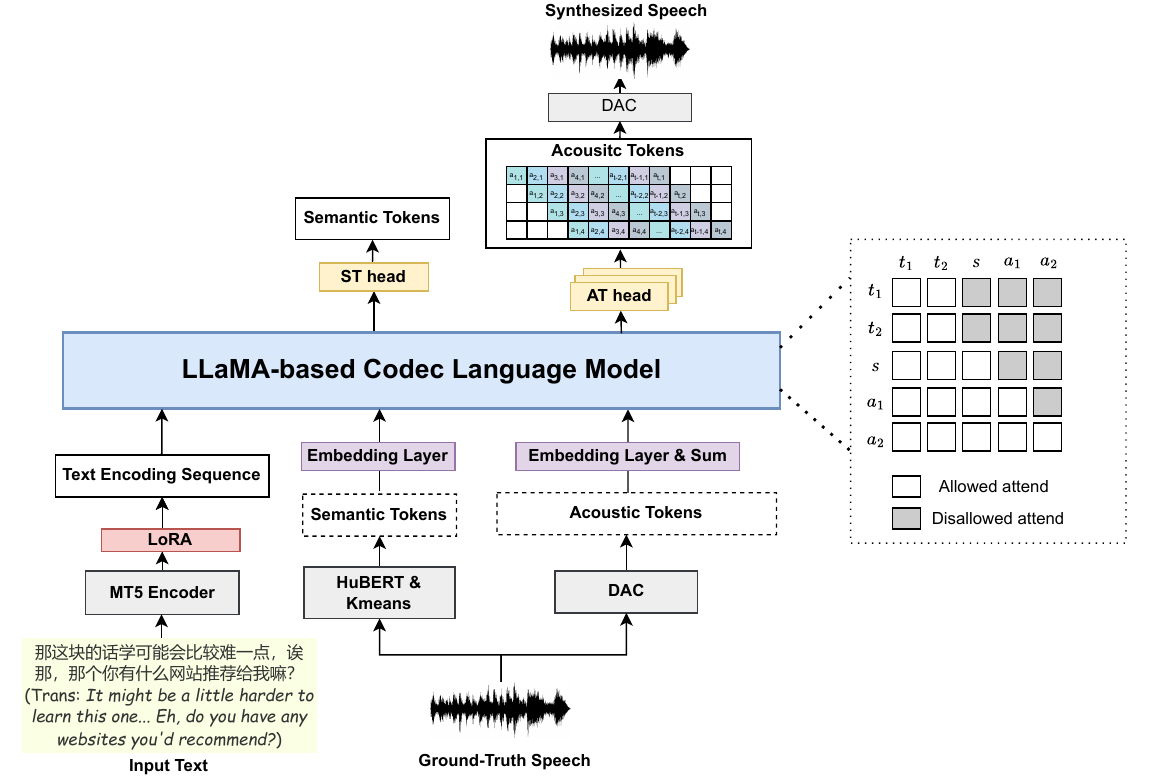}
  \caption{The model architecture in training procedure.}
  \label{fig:model}
  \vspace{-15pt}
\end{figure*}

\section{Data Preparation}
The system we have submitted adheres to the constrained track, meaning our model was trained exclusively using the WenetSpeech4TTS, MAGICDATA-RAMC, and HQ-Conversations datasets provided for this Challenge.
The duration and basic information of these datasets are presented in Table~\ref{tab:dataset}.
Subsequently, we will provide a comprehensive description of these datasets and the data processing steps we have undertaken.


WenetSpeech4TTS\footnote{\urlstyle{same}\url{https://huggingface.co/datasets/Wenetspeech4TTS/WenetSpeech4TTS}}\cite{wenetspeech4tts}, a multi-domain Mandarin corpus, is derived from the open-sourced WenetSpeech dataset \cite{zhang2022wenetspeech} and tailored for TTS tasks. 
The dataset is then divided into several subsets including Premium, Standard, Basic, and Rest based on data quality.
Nevertheless, our observations have revealed issues such as poor audio quality, low volume, speaker overlaps, and background noise in the Standard, Basic, and Rest subsets. 
Conversely, the Premium subset, comprising approximately 1000 hours of data, maintains relatively high quality. 

MAGICDATA-RAMC\footnote{\urlstyle{same}\url{https://www.openslr.org/123/}} is a conversational speech dataset featuring 663 speakers.
Due to the issues such as recording environment, background laughter, noise, and overlapping speech, the dataset’s overall quality is affected, making it less suitable for direct use in high-quality TTS models.
To minimize the impact of noise on the quality of synthesized speech within the MAGICDATA-RAMC dataset, we implemented a data preprocessing procedure. Specifically, utilizing the provided transcriptions, we segmented the sentences and meticulously eliminated segments containing overlapping speech, unintelligible noises, musical sounds, and laughter.

HQ-Conversations\footnote{\urlstyle{same}\url{https://www.magicdatatech.com/iscslp-2024}} is a high-quality dataset that includes 200 speakers and 100 hours of conversation. 
The predominant portion of this dataset exhibits a spontaneous style, aligning well with the characteristics of real human speech.

\begin{table}[th]\footnotesize
\renewcommand{\arraystretch}{1.0}
  \caption{The duration and basic information of datasets.}
  \label{tab:dataset}
  \centering
  \begin{tabular}{l|c|c} 
    \toprule
    \textbf{Datasets} &\textbf{Duration}  &\textbf{Basic Information} \\
    \midrule
    WenetSpeech4TTS &12800h& multi-domain Mandarin \\&& speech dataset for TTS~~~ \\
    \hline
    MAGICDATA-RAMC &180h& Natural Conversation \\&& speech Dataset~~~ \\
    \hline
    HQ-Conversations &100h& High-quality Spontaneous \\&& Conversation Dataset~~~ \\
    \bottomrule
  \end{tabular}
\end{table}

Reflecting on the analysis, during the pre-training stage, we utilized all available datasets, including the WenetSpeech4TTS, the MAGICDATA-RAMC and HQ-Conversations datasets, to equip the model with foundational zero-shot capabilities. 
In the fine-tuning stage, we used only the Premium subset of WenetSpeech4TTS and the HQ-Conversations dataset.
These two high-quality datasets were essential for improving the stability of the synthesis results and enhancing the model's ability to synthesize spontaneous-style speech.

\section{System}
In this part we will describe our Conversational Voice Clone System in detail. Firstly, we will overview the text representations and the discrete speech tokens, and then introduce the LLaMA-based codec language model with a delay pattern. Finally, we will introduce the classifier-free guidance for codec language model.
\vspace{-3pt}
\subsection{Feature Representation}

For the text representations, we utilize a pre-trained text encoder to convert raw text into a sub-word-level text encoding sequence. We choose the Multilingual T5 base model (MT5-base\footnote{\urlstyle{same}\url{https://huggingface.co/google/mt5-base}})\cite{xue2021mt5}, and enhance its pre-trained text encoder with trainable low-rank adaptation (LoRA) adapters. 
The raw text 
is passed to the MT5 encoder to generate the text encoding sequence $\textit{E}_{txt}=\{e_1, e_2, ..., e_m\}$, where $m$ is the number of subwords after text tokenization.

For speech semantic tokens (ST), we apply the self-supervised representation model Hubert\footnote{\urlstyle{same}\url{https://github.com/facebookresearch/fairseq/tree/main/examples/hubert}}\cite{hsu_hubert_2021} and K-means clustering to discretize 9th-layer hidden states into a 50Hz semantic token sequence, using a codebook size of 500.
For speech acoustic tokens (AT),
Descript Audio Codec\footnote{{\urlstyle{same}\url{https://github.com/descriptinc/descript-audio-codec}}} (DAC)\cite{kumar2024high} is used to extract the residual discrete acoustic tokens as intermediate representations $\textit{A}^{T*K}$, where $K$ is the number of residual quantizers of each frame. 
The DAC encoder has a total downsampling rate of 320 and flattens 12 codebooks. 

\subsection{Model Architecture}

As shown in Figure~\ref{fig:model}, our model is a codec language model-based TTS, which aims to generate acoustic tokens conditioned on the text encoding sequence. 
For the model bacbone, we leverage a  powerful transformer architecture LLaMA \cite{touvron2023llama}. 
Besides, we use fast and memory-efficient flash attention \cite{dao2022flashattention} to replace the original attention module.

The LLaMA-based codec language model consists of two stages: text to semantic token generation and acoustic token generation.
Both stages are modeled in an auto-regressive manner.
At the first stage, 
the text encoding sequence $\textit{E}_{txt}$ is used as condition to predict the semantic token sequence.
Semantic tokens are expected to offer abstract representations of spoken content that are free from prosodic features (e.g., duration) and speaking style. 
As a result, consecutive duplicate tokens are eliminated, adhering to the procedure described in \cite{kharitonov2023speak}. 
Once the semantic token prediction process is complete, the model will generate a $<S_{eos}>$ token to signify the end of ST prediction. 
This stage can be formalized as follows:
\vspace{-8pt}
\begin{equation}
    P(S|\textit{E}_{txt};\theta) = \prod_{t=1}^{T'} p(S_{t}|\textit{E}_{txt}, S_{<t};\theta) 
    \label{eq1}
\end{equation}
\vspace{-8pt}

In the second stage, the text encoding sequence, semantic tokens, and acoustic tokens $A_p$ from a speech prompt are used as conditions to predict the acoustic tokens for synthesis. 
Inspired by MusicGen \cite{copet2024simple} and ParlerTTS \cite{lyth2024natural}, we adopt an efficient delay pattern to predict all these residual acoustic tokens.
Before feeding into the model,
we apply a one-step offset to each layer of the residual acoustic tokens relative to the preceding layer, as illustrated in Figure \ref{fig:model}. This arrangement enables the prediction of the $k$-th AT at time $t$ to be conditioned on the prediction of the $(k-1)$-th AT from the same time step, 
which is believed to be more suitable for modelling the rich variations inherent in spontaneous style speech.
At timestep $t$, the model predicts all $K$ tokens simultaneously, using $K$ 
linear heads to project the hidden state of the final transformer block to $K$ sets of logits.
The second stage can be defined as:
\begin{equation}
\resizebox{1\hsize}{!}{
    $P(A\mid\textit{E}_{txt},S, A_p;\theta) = \prod_{t=1}^{T}\prod_{k=1}^{K} P(A_{(t-k+1,k)}\mid\textit{E}_{txt},S, A_p, A_{(<t, :)};\theta)$
}
    \label{eq2}
\end{equation}

The training loss is defined as the negative log likelihood in Equations \eqref{eq1} and \eqref{eq2}.
Because of the first layer of  residual acoustic tokens expressing more information than the latter layers, we introduce the $\alpha$ hyperparameters to adjust the weights and calculate the prediction loss on all acoustic tokens as the final loss.
\begin{equation}
\resizebox{0.9\hsize}{!}{
$\begin{aligned}
    \mathcal{L}(\theta) = -\log P(S|\textit{E}_{\text{txt}};\theta) -\log P(A|\textit{E}_{\text{txt}},S, A_p;\theta) \\
    = \mathcal{L}(S;\theta) + \sum_{k=1}^{K} \alpha_{k}\mathcal{L}_{k}(A_{(:,k)};\theta)
    \end{aligned}$
}
\end{equation}


\subsection{Classifier-Free Guidance for Codec Language Model}
Recent advancements in text generation have shown that CFG can improve coherence and alignment within LLMs \cite{sanchez2023stay}. 
Inspired by the success of CFG in speech generation model VoxInstruct \cite{zhou2024voxinstruct}, we also introduce CFG into our LLaMA-based codec language model to enhance the intelligibility of speech.

During training, the conditions in Equation \eqref{eq1} and \eqref{eq2} are substituted with an empty prompt at a certain probability.
When predicting semantic tokens, we mask the text encoding sequence, and when predicting acoustic tokens, 
we either mask text encoding sequence or the semantic token sequence.
Consequently, we can sample the token within the logits space by integrating both unconditional and conditional guidance at the inference stage.
The inference procedures are given by Equations \eqref{eq3}, \eqref{eq4} and \eqref{eq5}, where $\gamma$, 
$\alpha$ and $\beta$ denote guidance strength. 
The strength value is usually set to be over 1 for enhancing conditional guidance of LM generation. 
\begin{equation}
\resizebox{1\hsize}{!}{
     $log\hat{P}(S_{t}\mid\textit{E}_{txt}, S_{<t}) = \gamma log P(S_{t}\mid\textit{E}_{txt}, S_{<t}) + (1-\gamma) log P(S_{t}\mid\emptyset, S_{<t})$
    \label{eq3}
}
\end{equation}
\begin{equation}
\resizebox{1\hsize}{!}{
$\begin{aligned}
      log& \hat{P}(A_{t}|\textit{E}_{txt}, S, A_p, A_{<t}) \\ &= \alpha log P(A_{t}|\textit{E}_{txt}, S, A_p, A_{<t}) + (1-\alpha) log P(A_{t}|\emptyset,S, A_p, A_{<t})
\end{aligned}$
}
\label{eq4}
\end{equation}
\begin{equation}
\resizebox{1\hsize}{!}{
$\begin{aligned}
     log&\hat{P'}(A_{t}|\textit{E}_{txt},S, A_p, A_{<t})\\& = \beta log \hat{P}(A_{t}|\textit{E}_{txt}, S, A_p, A_{<t}) + (1-\beta) log P(A_{t}|\textit{E}_{txt}, \emptyset, A_p, A_{<t})
\end{aligned}$
}
\label{eq5}
\end{equation}
\vspace{-13pt}
\section{Experiments}
\subsection{Model Configurations}

We use the open-source models and parameters of MT5-base, DAC and HuBERT, with both DAC and HuBERT configured for a sampling rate of 16 kHz. 
The MT5-base text encoder comprises 12 transformer blocks with a hidden size of  768. 
We insert trainable LoRA adapters with parameters $\alpha=16$ and $r=16$ into query and value layers of this pre-trained text encoder. 
The LLaMA-based codec language model consists of 12 layer transformer blocks with 1024 hidden units and a feed-forward network dimension of 4096. 
The maximum sequence length is set to 2560, where the text encoding sequence occupies 512, and the ST and AT sequences occupy 2048. Padding is used to fill the shortfall of text encoding sequence part.
The loss weights $\alpha_k$ of the acoustic tokens are set to \{5, 2, 1, 0.5, 0.5, 0.2, 0.2, 0.2, 0.1, 0.1, 0.1, 0.1\} for 12 DAC layers.

\subsection{Training and Inference Details}
The codec language model with LoRA adapters was trained on 8 NVIDIA A100 GPUs, while all the other modules were frozen. 
Initially, we pre-trained the model using the large-scale WenetSpeech4TTS dataset for 450k iterations with a batch size of 64, employing a gradually decaying learning rate starting from 1e-4. 
A warm-up strategy was applied during the first 10k iterations. 
Following this, the model continued pre-training for an additional 250k iterations on comprehensive datasets including WenetSpeech4TTS, MAGICDATA-RAMC, and HQ-Conversations. 
Finally, we selected high-quality data, specifically the Premium-part corpus from WenetSpeech4TTS and HQ-Conversations, to fine-tune the model for 70k iterations. 
Thus the model was trained for 770k iterations totally.

What’s more, to facilitate unconditional generation as part of CFG, we masked the entire text encoding sequence or semantic token sequence with a probability of 0.1 during training.

During inference, considering the difficulty of synthesizing too long sentences that unseen in training, we segmented the long sentences in the test set at punctuation marks, and ensured that the segments are at least 30 characters long. 
The clips of these segments were then concatenated with 100ms silence intervals to form the final synthesized speech. 
Additionally, we set the CFG strength values as \(\alpha = 1.3\), \(\beta = 1.5\), and \(\gamma = 1.5\) in inference.

\subsection{Results}

\subsubsection{Evaluation metrics}

Subjective evaluations are carried out using Mean Opinion Score (MOS) tests to gauge speech quality, naturalness, speaker similarity, and the spontaneous style of the synthesized speech. 
The scores of the MOS tests are then averaged to determine the final ranking of all teams. 
Besides, the Character Error Rate (CER) and Speaker Encoder Cosine Similarity (SECS) are calculated as the objective evaluations, to measure the robustness and voice cloning performance of the systems.

\subsubsection{Evaluation results and discussion}

As displayed in Table \ref{tab:sub_res}, the subjective results show that our submitted system excels in speech naturalness and quality, with MOS scores of 3.80 and 3.84, ranking 1st and 2nd, respectively. 
In the realm of voice cloning, the model's performance is moderate, earning a MOS score of 3.49 and a 2nd place ranking. 
The spontaneous style MOS score, however, is somewhat lower at 3.33, which ranks 3rd.
Overall, our team achieves an average MOS of 3.61, securing the 3rd position among all teams.
For the objective evaluations, 
as shown in Table \ref{tab:obj_res}, the results reveal that our system has a CER of 10.29\%, ranking 2nd, 
and a SECS of 0.797, ranking 4th. 
This indicates that the proposed model exhibits relatively strong robustness, though its voice cloning performance could be further improved.

\vspace{-5pt}

\begin{table}[ht]
  \caption{MOS results of our system in CoVoC Challenge 2024}
  \label{tab:sub_res}
  \centering
  \begin{tabular}{lccc}
    \toprule
                 & \textbf{Score (mean)}     & \textbf{Score (std)} & \textbf{Rank}         \\
    \midrule
    Naturalness                    & 3.80         & 0.11  &  1st                \\
    Quality                        & 3.84         & 0.16    & 2nd              \\
    Similarity                     & 3.49         & 0.12  & 2nd                \\
    Spontaneous                     & 3.33         & 0.12   & 3rd               \\
    \midrule
    Average  &  3.61 & - & 3rd \\
    
    \bottomrule
  \end{tabular}
\end{table}
\vspace{-10pt}
\begin{table}[ht]
  \caption{Objective results in CoVoC Challenge 2024}
  \label{tab:obj_res}
  \centering
  \begin{tabular}{lcc}
    \toprule
    \textbf{System}      & \textbf{CER(\%)}     & \textbf{SECS}         \\
    \midrule
    Proposed                    & 10.29         & 0.797                \\
    
    \bottomrule
  \end{tabular}
\end{table}


The results above show that our proposed model is capable of synthesizing speech with commendable naturalness and quality, and the synthesized speech exhibits robustness.
As for text representations, we introduce text encoding sequences from the pre-trained MT5 text encoder instead of traditional phoneme sequences. 
This allows the model to better capture semantic information from raw text and improve prosodic performance. What's more, the CFG strategy is adopted to enhance conditional guidance generation, strengthening the influence of text and semantic tokens on acoustic token generation. 
This also improves the intelligibility and quality of the speech.
To better learn the spontaneous style 
characteristics in speech, we fine-tuned the model using the HQ-Conversations dataset. 
Concurrently, to ensure the model’s original voice cloning capabilities and high-quality generation as much as possible, we supplemented it with the Premium-part of WenetSpeech4TTS.
However, this compromise still sacrifices some of the spontaneity of the generated speech and impacted the speaker similarity, resulting in a slightly lower score for MOS.
Furthermore, our model relies solely on a data-driven approach to model spontaneous styles, without introducing explicit spontaneous behavior labels like SponLMTTS \cite{li2024spontaneous}. 
We believe that this approach has the potential to further improve the performance of our model.

\subsection{Case Study}

\begin{figure}[h]
    \centering
    \subfigure[] {
    \label{fig:case_study:sub_a}
    \includegraphics[width=0.4\linewidth]{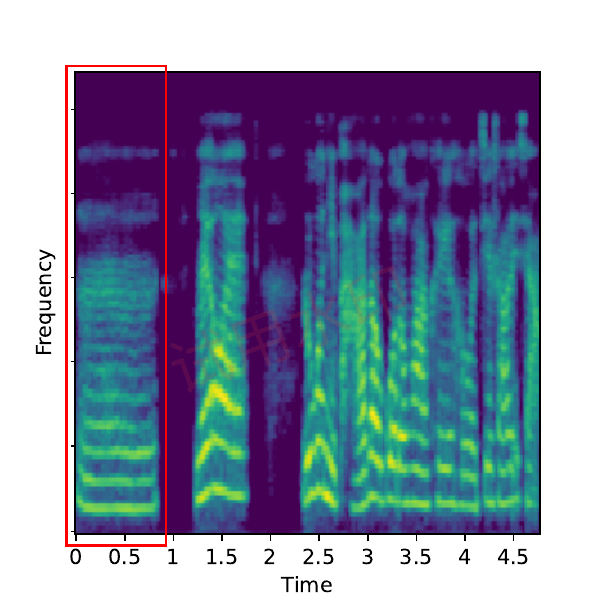}
    }
    \subfigure[] {
    \label{fig:case_study:sub_b}
    \includegraphics[width=0.4\linewidth]{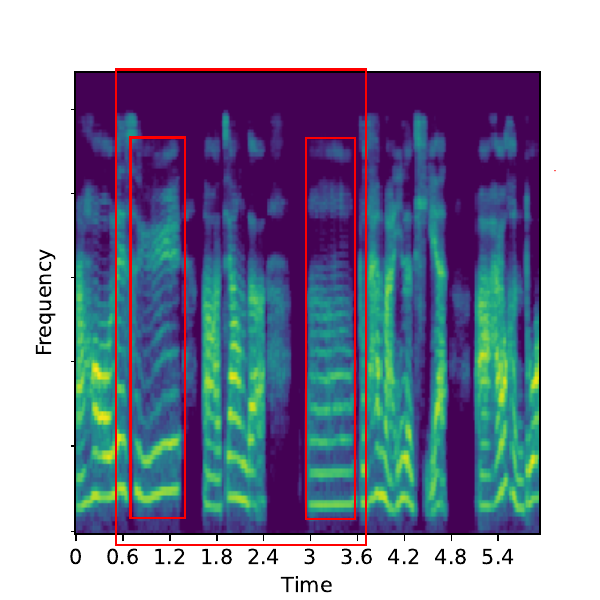}
    }
    \enspace\thinspace
    \caption{\begin{CJK*}{UTF8}{gbsn}The mel-spectrograms of speeches synthesized by proposed model. The texts of these two samples are ``嗯没有诶，如果你爬到过的话可以和我介绍一下'' (English translation: ``Um, no, I haven't. If you've climbed there before, you can tell me about it.'') and ``然后类似于啊这样的，嗯，不太满意的体验，啊还有很多。'' (English translation: ``And then, there were experiences like, uh, this, um, not very satisfying, uh, and there were many more.'') respectively. \end{CJK*}}
    \label{fig:case_study}
    \vspace{-15pt}
\end{figure}

\begin{CJK*}{UTF8}{gbsn}
To demonstrate the certain ability of our model 
in generating
spontaneous phenomena, we have visualized two samples to further analyze the detailed performance, as shown in Figure \ref{fig:case_study}. 
For the first sample in Figure \ref{fig:case_study:sub_a}, 
the duration of ``嗯'' (English translation: ``um''), as denoted by the red box, is distinctly longer than other words, indicating a spontaneous phenomenon where the speaker is thinking before speaking. 
For the second one in Figure \ref{fig:case_study:sub_b}, there is a noticeable hesitation between ``然后类似于'' (English translation: ``And then, there were experiences like'') and ``不太满意的体验'' (English translation: ``not very satisfying''), so the model accurately extends the duration of ``于'' (English translation: ``like'') in the left red box  and ``嗯'' (English translation: ``um'') in the right red box, showing the speaker is thinking about what to say next. 
The case study results show that our model can synthesize speech with spontaneous style and distinguish the pronunciation duration of different characters in the speech.
\end{CJK*}
\vspace{-10pt}

\section{Conclusions}

This paper introduces our zero-shot spontaneous style speech synthesis system for the CoVoC Challenge 2024. 
We propose a LLaMA-based codec language model with a delay pattern for spontaneous style voice cloning. We use CFG to enhance conditional guidance on token prediction, increasing the intelligibility of the speech content.
We also adopt effective data preprocessing operations and fine-tuning strategies to improve speech quality and spontaneity. 
The proposed system is ranked 3rd in the constrained track, with 1st place in speech naturalness and notable quality and similarity results.

\section{Acknowledgements}

This work is supported by National Natural Science Foundation of China (62076144), Shenzhen Key Laboratory of next generation interactive media innovative technology (ZDSYS20210623092001004) and Shenzhen Science and Technology Program (WDZC20220816140515001).

\bibliographystyle{IEEEtran}

\bibliography{main}


\end{document}